\input{aipcheck}

\documentclass[final,numberedheadings]{aipproc}
\usepackage{amsmath,amssymb}
\usepackage[colorlinks=true]{hyperref}

\layoutstyle{8x11single}

\newcommand{\comment}[1]{}

\newcommand{\lr}[1]{ \left( #1 \right) }
\newcommand{\lrs}[1]{ \left[ #1 \right] }
\newcommand{\lrc}[1]{ \left\{ #1 \right\} }

\newcommand{\re}{ {\rm Re} \, }

\renewcommand{\Re}{ {\rm Re} \, }
\renewcommand{\Im}{ {\rm Im} \, }

\newcommand{\diag}[1]{ {\rm diag} \, \left( #1 \right) }

\newcommand{\dslash}[1]{ #1 \!\!\!/}

\newcommand{\Eu}{\eta_{\uparrow}}
\newcommand{\Ed}{\eta_{\downarrow}}

\begin{document}

\title{Surface states of massive Dirac fermions with separated Weyl nodes}


\author{P.~V.~Buividovich}{
  address={Institute of Theoretical Physics, University of Regensburg,\\ D-93053 Germany, Regensburg, Universit\"{a}tsstra{\ss}e 31,\\ e-mail:pavel.buividovich@physik.uni-regensburg.de}
}

\begin{abstract}
 We derive the spectra of surface states for massive Dirac Hamiltonians with either momentum or energy separation between the left- and right-handed Weyl nodes. Momentum separation between the Weyl nodes corresponds to the explicitly broken time-reversal symmetry and the energy separation - to broken parity. Such Hamiltonians provide the simplest model description of Weyl semimetals. We find that the only effect of the energy separation between the Weyl nodes is to decrease the Fermi velocity in the linear dispersion relation of the surface states of massive Dirac Hamiltonian. In the case of broken time-reversal symmetry, the spectrum of surface states interpolates in a nontrivial way between the Fermi arc-type and the Dirac cone-type dispersion relations. In particular we find that for all values of the mass and the momentum separation between the Weyl nodes the surface states only exist in a strip of finite width in momentum space. We give also some simpler examples of surface states in order to make these notes more pedagogical.
\end{abstract}

\maketitle

\section{Introduction}
\label{sec:intro}

 Surface states play an extremely important role in the physics of topologically nontrivial states of matter, such as the topological insulators \cite{Qi:08:1} and the Weyl semimetals \cite{Wan:11:1,Burkov:11:1}. E.g. the nonzero conductivity of the topological insulators is saturated by topologically protected gapless excitations. The hallmark of the Weyl semimetal phase is the emergence of the Fermi arc in the spectrum of surface states, which is an open line of zero energy surface states joining the projections of the bulk Weyl nodes onto the boundary Brillouin zone. A beautiful argument relating 3D Weyl semimetals and 2D $Z_2$ topological insulators shows that Fermi arcs are also topologically protected \cite{Wan:11:1}.

 The aim of these notes is to provide a simple (and hopefully pedagogical) derivation of the spectrum of surface states within the low-energy effective model of topological insulators and Weyl semimetals, which is nothing but the continuum Dirac Hamiltonian. In Section (\ref{sec:surface_states_general}) we start with a general recipe for computing the spectrum of surface states, introducing the wave function of an ideal insulator as the boundary condition for the wave functions of any surface state. We then consider the simplest example of the surface states of a 3D $Z_2$ topological insulator, modelled by a Dirac Hamiltonian with a negative mass term.

 In Section \ref{sec:bs_muA} we consider the spectrum of surface states of a massive Dirac Hamiltonian with Weyl nodes which have different energies. In this case the surface states only exist if the Dirac mass term is negative, as for $Z_2$ topological insulators. The dispersion relation of the surface states is simply the isotropic Dirac cone with the Fermi velocity which decreases as the energy separation between the Weyl nodes grows.

 In Section \ref{sec:bs_B} we study surface states of a Dirac Hamiltonian with the momentum separation between the Weyl nodes. In Subsection \ref{subsec:FermiArcs_massless} we consider the simplest case of massless Dirac fermions and demonstrate the existence of the Fermi arc in the surface state spectrum. After that we consider massive Dirac fermions and find that the surface states interpolate in a nontrivial way between the Fermi arc-like and the Dirac cone-like dispersion relations. In particular, surface states exist only in a finite range of momenta in the surface Brillouin zone for all nonzero values of the Dirac mass and the momentum separating the Weyl nodes. When the momentum separation between the Weyl nodes is equal to the Dirac mass, the spectrum of surface states becomes purely one-dimensional.

\section{General recipe for the calculation of the spectrum of surface states}
\label{sec:surface_states_general}

\subsection{Surface states of a Dirac Hamiltonian in the presence of a flat boundary}
\label{subsec:dirac_surf_states_general}

 In these notes we will consider the single-particle three-dimensional Dirac Hamiltonians of the following form
\begin{eqnarray}
\label{DiracHamiltonianGeneral}
 h = -i \alpha_i \partial_i + \Phi\lr{\vec{x}}  ,
\end{eqnarray}
where $\alpha_i = -i \gamma_0 \gamma_i = \diag{\sigma_i, -\sigma_i}$ are the Dirac $\alpha$-matrices, $i = x, y, z$ label spatial coordinates and $\Phi\lr{\vec{x}}$ is some Hermitian matrix with two spinor indices which might in general depend on the coordinates. For example, the Dirac mass term corresponds to $\Phi = m \gamma_0$. The Dirac Hamiltonians of such a general form provide a reasonably accurate low-energy description of 3D $Z_2$ topological insulators and Weyl semimetals.

 In the following we will consider the Hamiltonian (\ref{DiracHamiltonianGeneral}) as a low-energy description of some crystal which hosts Dirac quasiparticles and which fills the infinite half-space $z > 0$ such that there is a flat boundary at $z = 0$ between the crystal and the vacuum. Moreover, we assume that at $z>0$ and $z<0$ the matrices $\Phi$ do not have any coordinate dependence, and denote $\Phi\lr{\vec{x}} \equiv \Phi_{>}$ for $z > 0$ and $\Phi\lr{\vec{x}} \equiv \Phi_{<}$ for $z < 0$. We are then interested in the surface states of the Hamiltonian (\ref{DiracHamiltonianGeneral}) confined to the plane $z = 0$, in a sense that their wave functions decay as $e^{\lambda_{>} z}$ for $z \rightarrow +\infty$ and as $e^{\lambda_{<} z}$ for $z \rightarrow -\infty$, with $\re \lambda_{>} < 0$ and $\re \lambda_{<} > 0$.

 For real crystals there should be also a factor of Fermi velocity $v_F$ in front of the derivative term in (\ref{DiracHamiltonianGeneral}). However, it can be absorbed into the rescaling of $\Phi\lr{\vec{x}}$ and the energy spectrum, and is not important for the following discussion.

 Since the system is translationally invariant in the $xy$ plane, we can partially diagonalize the Hamiltonian in the plane wave basis $e^{i k_x x + i k_y y}$. We then get the following equation for the eigenstates of $h$:
\begin{eqnarray}
\label{EigenstateEquationGeneral}
 \lr{\alpha_a k_a - i \alpha_z \partial_z + \Phi\lr{z}} \Psi\lr{k_a, z} = \epsilon\lr{k_a} \Psi\lr{k_a, z}
 \quad 
 \Rightarrow \quad \partial_z \Psi = i \alpha_z \lr{\epsilon - \alpha_a k_a - \Phi_{>}\theta\lr{z} - \Phi_{<}\theta\lr{-z}} \Psi  ,
\end{eqnarray}
where small Latin indices are used to label the two transverse coordinates: $a = x, y$. From the last equation above we immediately conclude that the coefficients $\lambda_{\lessgtr}$ which determine the decay of the wave function of the surface state away from the surface are the eigenvalues of the constant, space-independent $4 \times 4$ matrices
\begin{eqnarray}
\label{MMatrixDef}
 M_{\lessgtr} = i \alpha_z \lr{\epsilon - \alpha_a k_a - \Phi_{\lessgtr}} .
\end{eqnarray}
The corresponding eigenvectors are the Dirac spinors $\psi_{\lessgtr}$, for which we assume the normalization $|\psi_{\lessgtr}| = 1$. The solution of the equation (\ref{EigenstateEquationGeneral}) can be then written in the following general form:
\begin{eqnarray}
\label{BoundaryStateGeneralWF}
 \Psi\lr{z}
 =
 \mathcal{N}_{>} \psi_{>} e^{\lambda_{>} z} \theta\lr{z}
 +
 \mathcal{N}_{<} \psi_{<} e^{\lambda_{<} z} \theta\lr{-z}  ,
\end{eqnarray}
The normalization condition for the wave function (\ref{BoundaryStateGeneralWF}) reads
\begin{eqnarray}
\label{BoundaryStateGeneralWFNormalization}
 || \Psi ||^2
 = 
 |\mathcal{N}_{>}|^2 \int\limits_{-\infty}^{0} dz \, e^{2 \re \lambda_{>} z}
 +
 |\mathcal{N}_{<}|^2 \int\limits_{0}^{+\infty} dz \, e^{2 \re \lambda_{<} z}
 = 
 \frac{|\mathcal{N}_{>}|^2}{2 |\re \lambda_{>}|} + \frac{|\mathcal{N}_{<}|^2}{2 |\re \lambda_{<}|} = 1 .
\end{eqnarray}
In addition, in order to satisfy the equation (\ref{EigenstateEquationGeneral}) we have to require the continuity of the wave function across the boundary at $z = 0$, which gives the condition $\mathcal{N}_{>} \psi_{>} = \mathcal{N}_{<} \psi_{<}$. Taking into account that $\psi_{\lessgtr}$ are by definition normalized to unity, we conclude that the absolute values of $\mathcal{N}_{>}$ and $\mathcal{N}_{<}$ should be equal, and that the spinors $\psi_{>}$ and $\psi_{<}$ should be equal up to a phase: $\psi_{>} = n \psi_{<}$ with $|n|^2 = 1$.

 We thus conclude that the surface states exist if one can find such $\epsilon$ that the two matrices $M_{\lessgtr}$ have matching eigenvectors $\psi_{>} = n \psi_{<}$ for which the corresponding eigenvalues $\lambda_{\lessgtr}$ satisfy $\re \lambda_{>} < 0$, $\re \lambda_{<} > 0$.

\subsection{Wave function of an ideal insulator}
\label{subsec:ideal_insulator_wf}

 In practice, one is often interested in the surface states which live on the boundary between some topologically nontrivial material and the vacuum, which can be thought of as an ``ideal'' insulator. Since we know that in the vacuum the electrons are described by the Dirac Hamiltonian with a huge (as compared to typical scales in condensed matter physics) mass of $m_e \approx 0.5 \, {\rm MeV}$, we can model the vacuum electron Hamiltonian as the Dirac Hamiltonian (\ref{DiracHamiltonianGeneral}) with $\Phi = m_e \gamma_0$, assuming that the typical electron momenta and energies are much less than $m_e$. Strictly speaking, we should also take into account that inside the crystal the electron moves with a Fermi velocity $v_F$, and outside - with the speed of light. At the level of the eigenstate equations (\ref{EigenstateEquationGeneral}), however, the difference of the Fermi velocities will only result in a rescaling of $m_e$ by $v_F$, which will still be very large as compared to other scales in the problem.

 We will therefore assume that at $z < 0$ we have a vacuum, and calculate the eigenstate of the corresponding matrix $M_{<}$ in (\ref{MMatrixDef}) with positive real part of the eigenvalue, assuming $|k| \ll m_e$, $|\epsilon| \ll m_e$. Writing the matrix $M_{<}$ and its eigenstate $\psi_{<} = \lrc{\phi_{<}, \chi_{<}}$ in the chiral block form, we obtain the following eigenstate equation:
\begin{eqnarray}
\label{IIEigenEquationBlock1}
 i
 \left(
   \begin{array}{cc}
     \sigma_z & 0 \\
     0 & -\sigma_z \\
   \end{array}
 \right)
 \left(
   \begin{array}{cc}
     \epsilon - \dslash{k} & -m_e \\
     -m_e & \epsilon + \dslash{k} \\
   \end{array}
 \right)
 \left(
   \begin{array}{cc}
     \phi_{<} \\
     \chi_{<} \\
   \end{array}
 \right)
 = \lambda_{<}
 \left(
   \begin{array}{cc}
     \phi_{<} \\
     \chi_{<} \\
   \end{array}
 \right) ,
\end{eqnarray}
or, individually for each of the chiral components $\phi_{<}$, $\chi_{<}$:
\begin{eqnarray}
\label{IIEigenEquationBlock2}
 \chi_{<} = \frac{\epsilon - \dslash{k} + i \lambda_{<} \sigma_z}{m_e} \, \phi_{<}
 , \quad 
 \phi_{<} = \frac{\epsilon + \dslash{k} - i \lambda_{<} \sigma_z}{m_e} \, \chi_{<} ,
\end{eqnarray}
where we have denoted $\dslash{k} \equiv \sigma_x k_x + \sigma_y k_y$.
Combining the two above equations, we can express $\lambda_{<}$ in terms of $\epsilon$ as $\lambda_{<} = \sqrt{m_e^2 + k^2 - \epsilon^2}$. We have taken into account that the real part of $\lambda_{<}$ should be positive and denoted $k = \sqrt{k_x^2 + k_y^2}$. From (\ref{IIEigenEquationBlock2}) one can also read off the general form of the eigenstates of $M_{<}$:
\begin{eqnarray}
\label{IIEigenstate1}
 \psi_{<}\lr{k_a} = \mathcal{N} \, \left(
   \begin{array}{cc}
     \eta\lr{k_a} \\
     \frac{\epsilon - \dslash{k} + i \lambda_{<} \sigma_z}{m_e} \, \eta\lr{k_a} \\
   \end{array}
 \right) ,
\end{eqnarray}
where $\mathcal{N}$ is the normalization factor and $\eta\lr{k_a}$ is an arbitrary normalized two-component Weyl spinor which can have some momentum dependence.

 Taking into account the smallness of $\epsilon$ and $k$ as compared to $m_e$, we can write $\lambda_{<} \approx m_e$. We can also neglect $\epsilon$ and $\dslash{k}$ in the second component of (\ref{IIEigenEquationBlock2}). We conclude therefore that in the limit $k \ll m_e, |\epsilon| \ll m_e$ the eigenstates $\psi_{<}$ of $M_{<}$ have the following form:
\begin{eqnarray}
\label{IdealInsulatorEigenstate}
 \psi_{<}\lr{k_a} = \frac{1}{\sqrt{2}} \, \left(
   \begin{array}{cc}
     \eta\lr{k_a} \\
     i \sigma_z \, \eta\lr{k_a} \\
   \end{array}
 \right) .
\end{eqnarray}
We call this spinor ``the wave function of an ideal insulator''. We see that its form is completely independent of the momentum $k$, the energy $\epsilon$ and the electron mass. By virtue of the continuity of the wave function at the boundary of material, we conclude that all the surface states of any material bounded by the vacuum should have this structure. The nontrivial properties of the surface states are then encoded in the nontrivial momentum dependence of the Weyl spinor $\eta\lr{k_a}$.

 From now on we will always assume that the nontrivial material occupies the half-space at $z > 0$, and the vacuum is at $z < 0$. To shorten the notation, from now on we will also omit the subscript $>$ for the quantities characterizing the wave functions at $z > 0$ as well as the surface momentum $k_a$ from the arguments of the wave functions.

\subsection{Example: surface states of a 3D topological insulator and the spin-momentum locking}
\label{subsec:bs_TI}

 In order to illustrate the application of the above formulae on the simplest possible example, in this Subsection we consider the surface states of a 3D $Z_2$ topological insulator, which at low energies can be modelled by the Dirac Hamiltonian with a negative mass term $m$ \cite{Qi:08:1}. The eigenstate equation for the matrix $M$ has the same form as in (\ref{IIEigenEquationBlock1}) with the replacement $m_e \rightarrow m$, and the corresponding eigenstates read
\begin{eqnarray}
\label{Eigenstate1_TI}
 \psi = \mathcal{N} \, \left(
   \begin{array}{cc}
     \eta \\
     \frac{\epsilon - \dslash{k} + i \lambda \sigma_z}{m} \, \eta \\
   \end{array}
 \right) ,
\end{eqnarray}
where $\lambda = -\sqrt{m^2 + k^2 - \epsilon^2}$ and $\mathcal{N}$ is some normalization constant. At the surface of the topological insulator this wave function should be equal to the wave function (\ref{IdealInsulatorEigenstate}) of an ideal insulator. These boundary conditions immediately lead to the following linear equation for $\eta$:
\begin{eqnarray}
\label{Eigenstate2_TI}
  \frac{\epsilon - \dslash{k} + i \lambda \sigma_z}{m} \, \eta = i \sigma_z \eta .
\end{eqnarray}
The consistency condition for these linear equations yield the equation from which one can find the energies $\epsilon$ of the surface states:
\begin{eqnarray}
\label{Eigenstate3_TI}
  \epsilon^2 = k^2 - \lr{\lambda - m}^2 \Rightarrow \lambda = m, \quad \epsilon = \pm k .
\end{eqnarray}
Note that for this solution the real part of $\lambda$ is only negative if the Dirac mass $m$ is negative, that is, if our material has a nontrivial $Z_2$ topological index with respect to the vacuum, which we have characterized by the ideal insulator wave function (\ref{IdealInsulatorEigenstate}). Of course only the relative sign of the Dirac mass in the vacuum and in the material is important, and the assumption that the Dirac mass in the vacuum is positive is merely a conventional choice \cite{Qi:08:1}. We thus see that the dispersion relation of the surface states of a 3D $Z_2$ topological insulator is simply the Dirac cone with unit Fermi velocity:
\begin{eqnarray}
\label{dirac_cone_2D}
 \epsilon = \pm k \equiv \pm \sqrt{k_x^2 + k_y^2} .
\end{eqnarray}

 Substituting now the expressions (\ref{Eigenstate3_TI}) into (\ref{Eigenstate2_TI}), we immediately see that the Weyl spinor $\eta$ is the eigenstate of the spin operator $\dslash{k}$: $\dslash{k} \eta = \epsilon \eta$. Since $\eta = \lrc{\eta_{\uparrow}, \eta_{\downarrow}}$ encodes the spin polarization of the surface states, we conclude that for 3D $Z_2$ topological insulators the spin of the surface states is always aligned with the momentum. This is the famous spin-momentum locking mechanism, which prevents backscattering of surface states at impurities which cannot flip the spin (e.g. the non-magnetic impurities). Indeed, the back-scattered wave should have opposite orientations of both the momentum and the spin, but the processes of spin flips are highly suppressed in the absence of magnetization.

 For completeness, let us also remark that in real 3D topological insulators the spin of the surface states is actually not aligned, but rather perpendicular to the momentum. This is simply the consequence of the fact that for real 3D $Z_2$ topological insulators the kinetic term in the effective Dirac Hamiltonian should be written as $\dslash{k} = k_x \sigma_y - k_y \sigma_x$ (see e.g. \cite{Fu:07:2,Fu:09:1,Vazifeh:13:1}). This specific permutation of $x$ and $y$ indices in the Dirac Hamiltonian is a direct reflection of a strong spin-orbital coupling, an important feature of all topological insulators. However, as long as one is interested only in the energy spectrum of the surface states, one can use the form $\dslash{k} = k_x \sigma_x + k_y \sigma_y$ which is common in high-energy physics and which only differs from $\dslash{k} = k_x \sigma_y - k_y \sigma_x$ by a redefinition of Pauli $\sigma$-matrices.

\section{Surface states for massive Dirac fermions with chirality imbalance}
\label{sec:bs_muA}

 Let us now consider a more nontrivial example of a massive Dirac Hamiltonian for which the left- and the right-handed Weyl nodes have different energies. Such Hamiltonian might serve as a low-energy effective theory of a Weyl semimetal with broken parity \cite{Vazifeh:13:1,Goswami:12:1,Goswami:13:1,Balents:12:1}. In particular, such a material should support the Chiral Magnetic Effect \cite{Kharzeev:08:2,Goswami:13:1}. Without loss of generality, we can assume that the energies of the Weyl nodes are $\pm \mu_A$, where $\mu_A$ is the chiral chemical potential \cite{Kharzeev:08:2}.

 We now consider the case when the space at $z < 0$ is an ideal insulator (see the above section), and the space at $z > 0$ is characterized by some mass term $m$ and the chiral chemical potential $\mu_A$. In this case the bulk Hamiltonian in momentum space and the corresponding matrix $\Phi$ in (\ref{DiracHamiltonianGeneral}) have the form
\begin{eqnarray}
\label{PhiPlus_muA}
 h = \left(
   \begin{array}{cc}
     k_i \sigma_i - \mu_A & m \\
     m & -\lr{k_i \sigma_i - \mu_A} \\
   \end{array}
 \right),
 \quad
 \Phi = \left(
   \begin{array}{cc}
     -\mu_A & m \\
     m & \mu_A \\
   \end{array}
 \right) .
\end{eqnarray}
The bulk energy spectrum which corresponds to such a Dirac Hamiltonian is
\begin{eqnarray}
\label{bulk_spectrum_muA}
 E_{s,\sigma}\lr{\vec{k}} = s \sqrt{\lr{|\vec{k}| - \sigma \mu_A}^2 + m^2} , \quad s, \sigma = \pm 1 .
\end{eqnarray}
The eigenvalue equation for the matrix $M$ in (\ref{MMatrixDef}) in this case has the form
\begin{eqnarray}
\label{EigenstateEquation_muA}
 \left(
   \begin{array}{cc}
     \epsilon - \dslash{k} + \mu_A & -m \\
     -m & \epsilon + \dslash{k} - \mu_A \\
   \end{array}
 \right)
 \left(
   \begin{array}{c}
     \phi \\
     \chi \\
   \end{array}
 \right)
 =
 \left(
   \begin{array}{c}
     -i \sigma_z \lambda \phi \\
      i \sigma_z \lambda \chi \\
   \end{array}
 \right) ,
\end{eqnarray}
or, in a component-wise form
\begin{eqnarray}
\label{EigenstateEquation_muA_components}
 \chi = \frac{\epsilon - \dslash{k} + \mu_A + i \sigma_z \lambda}{m} \, \phi ,
 \quad 
 \phi = \frac{\epsilon + \dslash{k} - \mu_A - i \sigma_z \lambda}{m} \, \chi .
\end{eqnarray}
For simplicity, we assume now that we are interested in the low-energy excitations with $\epsilon < m$. Then the compatibility condition for the equations (\ref{EigenstateEquation_muA_components}) leads to the following two values of $\lambda$ with negative real parts:
\begin{eqnarray}
\label{lambda_muA}
 \lambda_{\sigma} = - \sqrt{k^2 + \lr{\sqrt{m^2 - \epsilon^2} + i \sigma \mu_A}^2} ,
\end{eqnarray}
where we assume that $\sqrt{m^2 - \epsilon^2} > 0$. We see that in the presence of the chiral chemical potential $\mu_A$ $\lambda_{\sigma}$ acquire some nonzero imaginary part. Moreover, $\lambda_{+}$ and $\lambda_{-}$ are now complex conjugate. Remembering that the wave functions of the surface states depend on the ``depth'' coordinate $z$ as $e^{\lambda z}$, we conclude that now the wave functions of the surface states should exhibit some oscillations as they decay at large $z$.

 By a direct substitution one can check that $\phi$ and $\chi$ should be proportional to the eigenvectors $\eta_{\sigma}$ of the operator $\dslash{k} - i \sigma_z \lambda_{\sigma}$:
\begin{eqnarray}
\label{eta_def_muA}
 \lr{\dslash{k} - i \sigma_z \lambda_{\sigma}} \eta_{\sigma} = -i \sigma \sqrt{\lambda_{\sigma}^2 - k^2} \eta_{\sigma}
= 
-i \sigma \lr{\sqrt{m^2 - \epsilon^2} + i \sigma \mu_A} ,
\end{eqnarray}
where in the last line we have used the identity $\lambda_{\sigma}^2 - k^2 = \lr{\sqrt{m^2 - \epsilon^2} + i \sigma \mu_A}^2$ which follows from (\ref{lambda_muA}). Up to normalization $\eta_{\sigma}$ can be written as \begin{eqnarray}
\label{eta_components_muA}
 \eta_{\sigma} = \lrc{1, \theta_{\sigma}},
 \quad 
\theta_{\sigma} = \frac{i \lambda_{\sigma} + \mu_A - i \sigma \sqrt{m^2 - \epsilon^2}}{k_x - i k_y} .
\end{eqnarray}
Using (\ref{EigenstateEquation_muA_components}), we can now find the eigenvectors of $M$ which correspond to the eigenvalues (\ref{lambda_muA}):
\begin{eqnarray}
\label{Psi_muA}
 \psi = \left( \begin{array}{c}
          \eta_{\sigma} \\
          \xi_{\sigma} \eta_{\sigma}
        \end{array} \right) ,
\quad 
 \xi_{\sigma} = \epsilon + \mu_A + i \sigma \lr{\sqrt{m^2 - \epsilon^2} + i \sigma \mu_A}
 = 
 \epsilon + i \sigma \sqrt{m^2 - \epsilon^2} .
\end{eqnarray}
Since $\psi_{\sigma}$, $\sigma = \pm 1$ are the two independent eigenstates of $M$ with negative real parts, the wave functions of surface states can be expressed as some linear combination of $\psi_{\sigma}$: $\Psi\lr{z} = \sum\limits_{\sigma = \pm} c_{\sigma} \psi_{\sigma} e^{\lambda_{\sigma} z}$. In order to satisfy the continuity of the wave function, we have to require that at $z = 0$ $\Psi\lr{z}$ is equal to the ideal insulator wave function (\ref{IdealInsulatorEigenstate}). This immediately leads to the following equations:
\begin{eqnarray}
\label{bs_muA_eq1}
 c_{+} \eta_{+} + c_{-} \eta_{-} = \eta, \quad
 c_{+} \xi_{+} \eta_{+} + c_{-} \xi_{-} \eta_{-} = i \sigma_z \eta ,
 \, \Rightarrow \,
 c_{+} \lr{\xi_{+} \eta_{+} - i \sigma_z \eta_{+}} + c_{-} \lr{\xi_{-} \eta_{-} - i \sigma_z \eta_{-}} = 0 .
\end{eqnarray}
Using now the explicit form of $\eta_{\sigma}$, we arrive at the following system of equations for $c_{\pm}$:
\begin{eqnarray}
\label{bs_muA_eq3}
 c_{+} \lr{\xi_{+} - i} + c_{-} \lr{\xi_{-} - i} = 0,
 \quad 
 c_{+} \theta_{+} \lr{\xi_{+} + i}  + c_{-} \theta_{-} \lr{\xi_{-} + i} = 0.
\end{eqnarray}
These equations are compatible if
\begin{eqnarray}
\label{bs_muA_eq4}
 \lr{\xi_{+} - i} \lr{\xi_{-} + i} \theta_{-} = \lr{\xi_{+} + i} \lr{\xi_{-} - i} \theta_{+} .
\end{eqnarray}
The real values of $\epsilon$ at which this equation is satisfied are the energies of the surface states. Using the explicit form of $\xi_{\sigma}$ from (\ref{Psi_muA}) and $\theta_{\sigma}$ from (\ref{eta_components_muA}), after some algebra we can rewrite the above equation as
\begin{eqnarray}
\label{bs_muA_eq5}
 \lr{1 + \frac{m}{\sqrt{m^2 - \epsilon^2}}} \lambda_{+} + \lr{1 - \frac{m}{\sqrt{m^2 - \epsilon^2}}} \lambda_{-}
= 
2 \lr{m + i \mu_A} .
\end{eqnarray}
Taking into account that $\lambda_{+}$ and $\lambda_{-}$ are complex conjugate and that $m$, $\epsilon$, $\sqrt{m^2 - \epsilon^2}$ and $\mu_A$ are real, one can easily see that the above equation is equivalent to
\begin{eqnarray}
\label{bs_muA_eq6}
 \Re \lambda_{+} = m, \quad \Im \lambda_{+} = \mu_A \frac{\sqrt{m^2 - \epsilon^2}}{m} .
\end{eqnarray}
From the first equation it becomes obvious again that the surface states only exist if $m < 0$. This means that surface states only exist if the Weyl semimetal is simultaneously also the $Z_2$ topological insulator. These features in fact do not contradict each other, since the dispersion relation (\ref{bulk_spectrum_muA}) features both the Weyl nodes at the energies $E = \pm \sqrt{\mu_A^2 + m^2}$ and the gap of size $2 |m|$ centered around $E = 0$. While the conventional ohmic conductivity vanishes at zero temperature because of this gap, the chiral magnetic conductivity is still nonzero due to the existence of Weyl nodes \cite{Ren:11:1,Buividovich:13:8}.

 In order to find the energies $\epsilon$ of surface states, we have now to solve the equation
\begin{eqnarray}
\label{bs_muA_eq7}
 - \sqrt{k^2 + \lr{\sqrt{m^2 - \epsilon^2} + i \mu_A}^2}
 = 
 m + i \mu_A \frac{\sqrt{m^2 - \epsilon^2}}{m} .
\end{eqnarray}
Squaring both the r.h.s. and the l.h.s. of this equation we finally arrive at the following dispersion relation for the surface states in the presence of chiral chemical potential:
\begin{eqnarray}
\label{bs_muA_final}
 \epsilon = \pm \frac{|k|}{\sqrt{1 + \frac{\mu_A^2}{m^2}}} .
\end{eqnarray}
We see that the dispersion relation at nonzero chiral chemical potential is still the Dirac cone, and the only effect of the chiral chemical potential is to decrease the Fermi velocity $v_F \sim \frac{m}{\sqrt{m^2 + \mu_A^2}}$. It is interesting to note that the Fermi velocity of the surface states appears to be different (and larger for $\mu_A < m$) than the bulk Fermi velocity $V_F$, which according to (\ref{bulk_spectrum_muA}) is equal to
\begin{eqnarray}
\label{muA_bulk_vF}
 V_F = \left| \frac{\partial}{\partial k} E_{s,\sigma}\lr{k}|_{k = 0} \right| = \frac{|\mu_A|}{\sqrt{\mu_A^2 + m^2}} .
\end{eqnarray}

\section{Surface states for massive Dirac fermions with momentum separation between the Weyl nodes}
\label{sec:bs_B}

 In this Section we consider the case of Dirac Hamiltonian with momentum separation between the Weyl nodes. Such a Hamiltonian provides a low-energy effective description of time-reversal breaking Weyl semimetals \cite{Wan:11:1,Vazifeh:13:1,Goswami:12:1}. Physically, momentum separation between the Weyl nodes can be achieved, for example, by magnetic doping of a 3D topological insulator \cite{Wan:11:1, Sekine:13:1}. Direct signatures of the momentum separation between the Weyl nodes are the anomalous Hall effect \cite{Vazifeh:13:1,Goswami:12:1} as well as the existence of the Fermi arc in the spectrum of surface states - an open line of topologically protected zero energy states which joins the projections of the bulk Weyl nodes onto the surface Brillouin zone \cite{Wan:11:1}.

 We first illustrate the emergence of the Fermi arc states in the simplest case of massless Dirac fermions with Weyl nodes at different momenta, and then consider the more general and complicated case of massive Dirac fermions. While similar calculations were presented in \cite{Goswami:12:1, Shovkovy:14:1}, here we extend the analysis of the surface states also to the case when absolute value of the Dirac mass is so large that the Weyl nodes no longer exist and explicitly follow the interpolation between the Fermi arc-like and the Dirac cone-like dispersion relations.

\subsection{Fermi arcs for massless Dirac fermions}
\label{subsec:FermiArcs_massless}

 For massless Dirac fermions with Weyl nodes at $\vec{k} = \pm \vec{b}$ the bulk Hamiltonian in the momentum space and the corresponding matrix $\Phi$ in (\ref{DiracHamiltonianGeneral}) have the form
\begin{eqnarray}
\label{PhiPlus_farcs}
 h = \left(
 \begin{array}{cc}
 \sigma_i \lr{k_i - b_i} & 0 \\
 0 & - \sigma_i \lr{k_i + b_i} \\
 \end{array}
 \right),
 \quad
 \Phi = \left(
 \begin{array}{cc}
  - \dslash{b} & 0 \\
 0 & - \dslash{b} \\
 \end{array}
 \right) ,
\end{eqnarray}
where we assume that $\vec{b} = \lrc{b, 0, 0}$. Fermi arcs appear in the spectrum in this case if the boundary of the Weyl semimetal is parallel to the $xy$ plane, that is, exactly in the case which we consider.

 With $\Phi$ given by (\ref{PhiPlus_farcs}), the equations for the eigenstates $\psi = \lrc{\phi, \chi}$ of $M$ can be written as
\begin{eqnarray}
\label{EigenstateEquation_farcs}
 \lr{\epsilon - \dslash{k} + \dslash{b} + i \sigma_z \lambda} \phi = 0,
 \quad 
 \lr{\epsilon + \dslash{k} + \dslash{b} - i \sigma_z \lambda} \chi = 0.
\end{eqnarray}
From these equations, we find two independent solutions with negative real $\lambda$:
\begin{eqnarray}
\label{Eigenstates_farcs}
 \lambda_1 = - \sqrt{|k - b|^2 - \epsilon^2},
 \quad 
 \phi_1 = \mathcal{N}_1 \lrc{1, \frac{\bar{k} - \bar{b}}{\epsilon - i \lambda_1}},
 \quad \chi_1 = 0 ,
 \nonumber \\
 \lambda_2 = - \sqrt{|k + b|^2 - \epsilon^2},
 \quad 
 \phi_2 = 0, \quad
 \chi_2 = \mathcal{N}_2 \lrc{1, - \frac{\bar{k} + \bar{b}}{\epsilon + i \lambda_2}} ,
\end{eqnarray}
where $\mathcal{N}_1$ and $\mathcal{N}_2$ are some normalization factors. Taking into account that for the vacuum boundary conditions (\ref{IdealInsulatorEigenstate}) $\lrc{\phi, \chi} = \mathcal{N} \lrc{\eta, i \sigma_z \eta}$, we can write the following equation for the general solution which should be representable as some linear combination of $\psi_1 = \lrc{\phi_1, \chi_1}$ and $\psi_2 = \lrc{\phi_2, \chi_2}$:
\begin{eqnarray}
\label{BC_equation_farcs}
 \left(
 \begin{array}{c}
 \mathcal{N}_1 \\
 \mathcal{N}_1 \frac{\bar{k} - \bar{b}}{\epsilon - i \lambda_1} \\
 \mathcal{N}_2 \\
 -\mathcal{N}_2\frac{\bar{k} + \bar{b}}{\epsilon + i \lambda_2} \\
 \end{array}
 \right)
 =
 \left(
 \begin{array}{c}
 \Eu \\
 \Ed \\
 i \Eu \\
 -i \Ed \\
 \end{array}
 \right) .
\end{eqnarray}
After some simple algebraic transformations we find the following equation for $\epsilon$:
\begin{eqnarray}
\label{EnergyEq1_farcs}
 \frac{k - b}{\epsilon + i \lambda_1} = \frac{k + b}{\epsilon - i \lambda_2} ,
\end{eqnarray}
or, in somewhat more explicit form
\begin{eqnarray}
\label{EnergyEq2_farcs}
 \frac{\epsilon - i \sqrt{|k - b|^2 - \epsilon^2}}{k - b} = \frac{\epsilon + i \sqrt{|k + b|^2 - \epsilon^2}}{k + b} .
\end{eqnarray}
It is now easy to guess the following solution to this equation:
\begin{eqnarray}
\label{Energy_farcs}
 \epsilon = - k_y, \quad |k_x|< |b| .
\end{eqnarray}
This is the Fermi arc solution. We see that there is a line of zero energy joining the projections of the bulk Weyl nodes onto the surface momentum space. The dispersion relation is effectively one-dimensional in the direction perpendicular to the separation between the Weyl nodes. Moreover, surface states only exist in a finite strip in momentum space with $|k_x| < |b|$.

\subsection{Surface states for massive Dirac fermions with broken time-reversal symmetry}
\label{subsec:fermi_arcs_massive}

 We now consider the more general case of massive Dirac Hamiltonian with the time-reversal breaking terms of the form (\ref{PhiPlus_farcs}). Correspondingly, the momentum-space bulk Hamiltonian and the matrix $\Phi$ in (\ref{DiracHamiltonianGeneral}) have the form
\begin{eqnarray}
\label{PhiPlus_B}
 h = \left(
 \begin{array}{cc}
 \sigma_i \lr{k_i - b_i} & m \\
 m & - \sigma_i \lr{k_i + b_i} \\
 \end{array}
 \right),
 \quad
 \Phi = \left(
   \begin{array}{cc}
     -\dslash{b} & m \\
     m & -\dslash{b} \\
   \end{array}
 \right) .
\end{eqnarray}
The bulk dispersion relation of this Hamiltonian is \cite{Goswami:12:1}:
\begin{eqnarray}
\label{WSMB_bulk_spectrum}
 E_{s, \sigma}\lr{\vec{k}} = s \sqrt{\lr{\sqrt{k_x^2 + m^2} + \sigma b}^2 + k_y^2 + k_z^2}, \quad s, \sigma = \pm 1 .
\end{eqnarray}
Physically, such Hamiltonian describes a $Z_2$ 3D topological insulator with magnetic doping which explicitly breaks time-reversal symmetry. It is important to note that the distance between the Weyl nodes is now $2 \sqrt{b^2 - m^2}$ rather than $2 b$. Thus the Dirac mass term tends to move the Weyl nodes together and eventually annihilates them if $|b| < |m|$. If the Dirac mass $m$ is negative, this situation corresponds to the 3D topological insulator for which the magnetic doping is still small compared to the topological mass gap $m$.

 Let us now study the surface states for such a Hamiltonian. Partly this has been done in \cite{Goswami:12:1}, but here we extend the analysis also to the case of large negative Dirac masses with $m < -|b|$ in order to see how the Dirac cone dispersion relation (\ref{dirac_cone_2D}) typical for the surface states of 3D topological insulators transforms into the Fermi arc-like dispersion relation (\ref{Energy_farcs}) typical for Weyl semimetals as the parameter $b$ is tuned across the critical value $m = -|b|$.

The eigenvalue equation for the matrix $M$ now has the form
\begin{eqnarray}
\label{EigenstateEquation_B}
 \left(
   \begin{array}{cc}
     \epsilon - \dslash{k} + \dslash{b} & -m \\
     -m & \epsilon + \dslash{k} + \dslash{b} \\
   \end{array}
 \right)
 \left(
   \begin{array}{c}
     \phi \\
     \chi \\
   \end{array}
 \right)
 =
 \left(
   \begin{array}{c}
     -i \sigma_z \lambda \phi \\
      i \sigma_z \lambda \chi \\
   \end{array}
 \right) ,
\end{eqnarray}
or, in a component-wise form
\begin{eqnarray}
\label{EigenstateEquation_B_components}
 \chi = \frac{\epsilon - \dslash{k} + \dslash{b} + i \sigma_z \lambda}{m} \, \phi ,
 \quad 
 \phi = \frac{\epsilon + \dslash{k} + \dslash{b} - i \sigma_z \lambda}{m} \, \chi .
\end{eqnarray}
Substituting the first equation into the second one, we obtain the following equation for $\phi$:
\begin{eqnarray}
\label{EigenstateEquation_B_phi1}
 \lr{\epsilon^2 + 2 \epsilon \dslash{b} + b^2 + \lrs{\dslash{k} - i \lambda \sigma_z, \dslash{b}} - k^2 + \lambda^2 - m^2} \phi = 0 .
\end{eqnarray}
We see thus that $\phi$ should be the eigenstate of the operator $\mathrm{D} = 2 \epsilon \dslash{b} + \lrs{\dslash{k}, \dslash{b}} - i \lambda \lrs{\sigma_z, \dslash{b}}$, which can be written as the following $2 \times 2$ matrix:
\begin{eqnarray}
\label{EigenstateEquation_B_phi2}
 \mathrm{D} = 2 b \left(
                        \begin{array}{cc}
                          -i k_y & \epsilon - i \lambda \\
                          \epsilon + i \lambda & i k_y \\
                        \end{array}
                      \right) .
\end{eqnarray}
The eigenvalues of this matrix are $\mathrm{d}_{\sigma} = 2 \sigma b_x \sqrt{\lambda^2 + \epsilon^2 - k_y^2}$, with $\sigma = \pm 1$. The corresponding eigenvectors have the following form, up to normalization:
\begin{eqnarray}
\label{phi_sigma_B}
 \phi_{\sigma} = \lrc{1, \theta_{\sigma}}, \quad \theta_{\sigma} = \frac{i k_y + \sigma \sqrt{\lambda^2 + \epsilon^2 - k_y^2}}{\epsilon - i \lambda}
 = 
 \frac{i k_y + \sigma \sqrt{k_x^2 + m^2} - b}{\epsilon - i \lambda} ,
\end{eqnarray}
where in the last line we have used the explicit expression for $\lambda$ in terms of $\sigma$ (\ref{lambda_B}), given below. Substituting this expression for $\phi_{\sigma}$ into the first equation (\ref{EigenstateEquation_B_components}), we also find $\chi_{\sigma}$:
\begin{eqnarray}
\label{chi_sigma_B}
 \chi_{\sigma} = \frac{1}{m} \,
 \left(
  \begin{array}{c}
   \lr{\epsilon + i \lambda} + \lr{-k_x + i k_y + b} \theta_{\sigma} \\
   \lr{-k_x - i k_y + b} + \lr{\epsilon - i \lambda} \theta_{\sigma}
  \end{array}
 \right)
\end{eqnarray}
Substituting the eigenstates of $\mathrm{D}_{\phi}$ into (\ref{EigenstateEquation_B_phi1}), we obtain the following equation for $\lambda$:
\begin{eqnarray}
\label{EigenstateEquation_B_phi3}
 \epsilon^2 + b^2 - k^2 + \lambda^2 - m^2 + 2 \sigma b \sqrt{\lambda^2 + \epsilon^2 - k_y^2} = 0 .
\end{eqnarray}
The solutions of this equation with negative real part are given by
\begin{eqnarray}
\label{lambda_B}
 \lambda_{\sigma} = - \sqrt{\lr{\sqrt{k_x^2 + m^2} - \sigma b}^2 + k_y^2 - \epsilon^2} .
\end{eqnarray}
The square root in the brackets (inside the outer square root) can in principle have an arbitrary sign. In order to match the expression for the component $\theta_{\sigma}$ of $\phi_{\sigma}$ given by the last line of (\ref{phi_sigma_B}), we have to set this sign to $+1$, assuming that $\sqrt{k_x^2 + m^2} > 0$. Now $\lambda_{\sigma}$ should be substituted in the expressions (\ref{phi_sigma_B}) and (\ref{chi_sigma_B}) above. It is obvious that $\lambda_{\sigma}$ given by (\ref{lambda_B}) have either zero imaginary or zero real part. In order to find the localized surface states, we should only consider the solutions with nonzero (and negative) real part.

\begin{ltxfigure}[htpb]
  \centering
  \includegraphics[width=5cm]{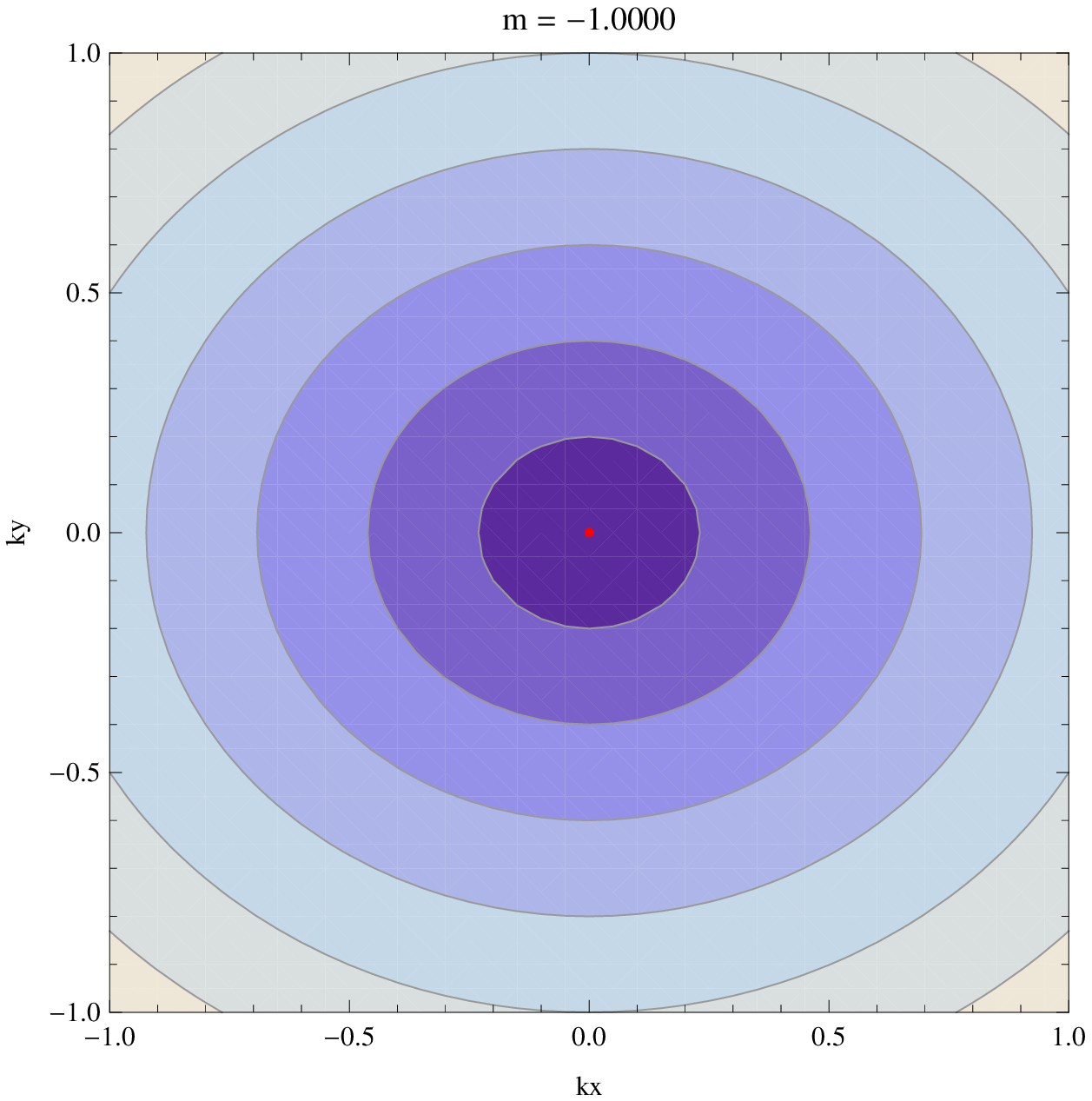}
  \includegraphics[width=5cm]{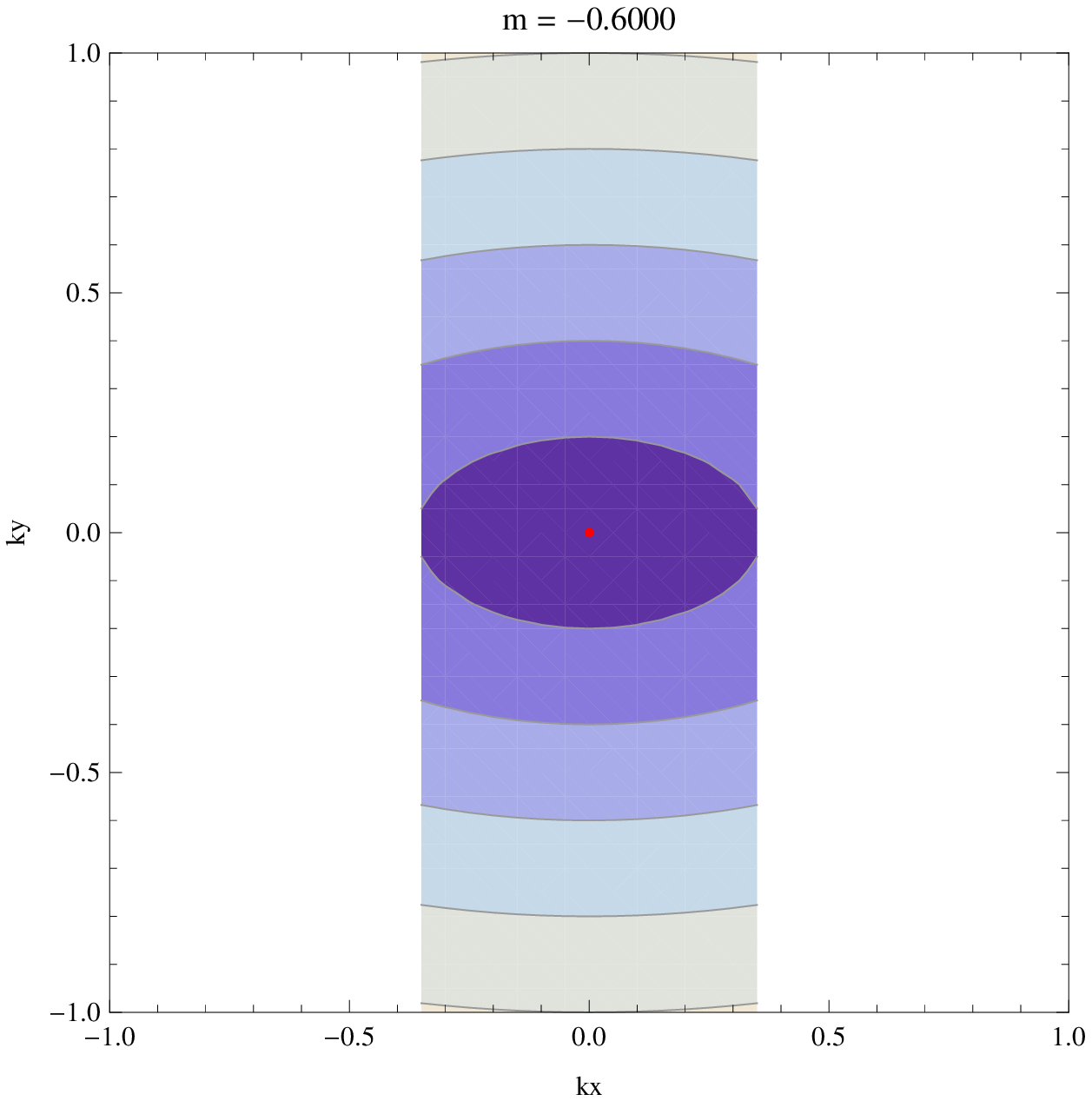}\\
  \includegraphics[width=5cm]{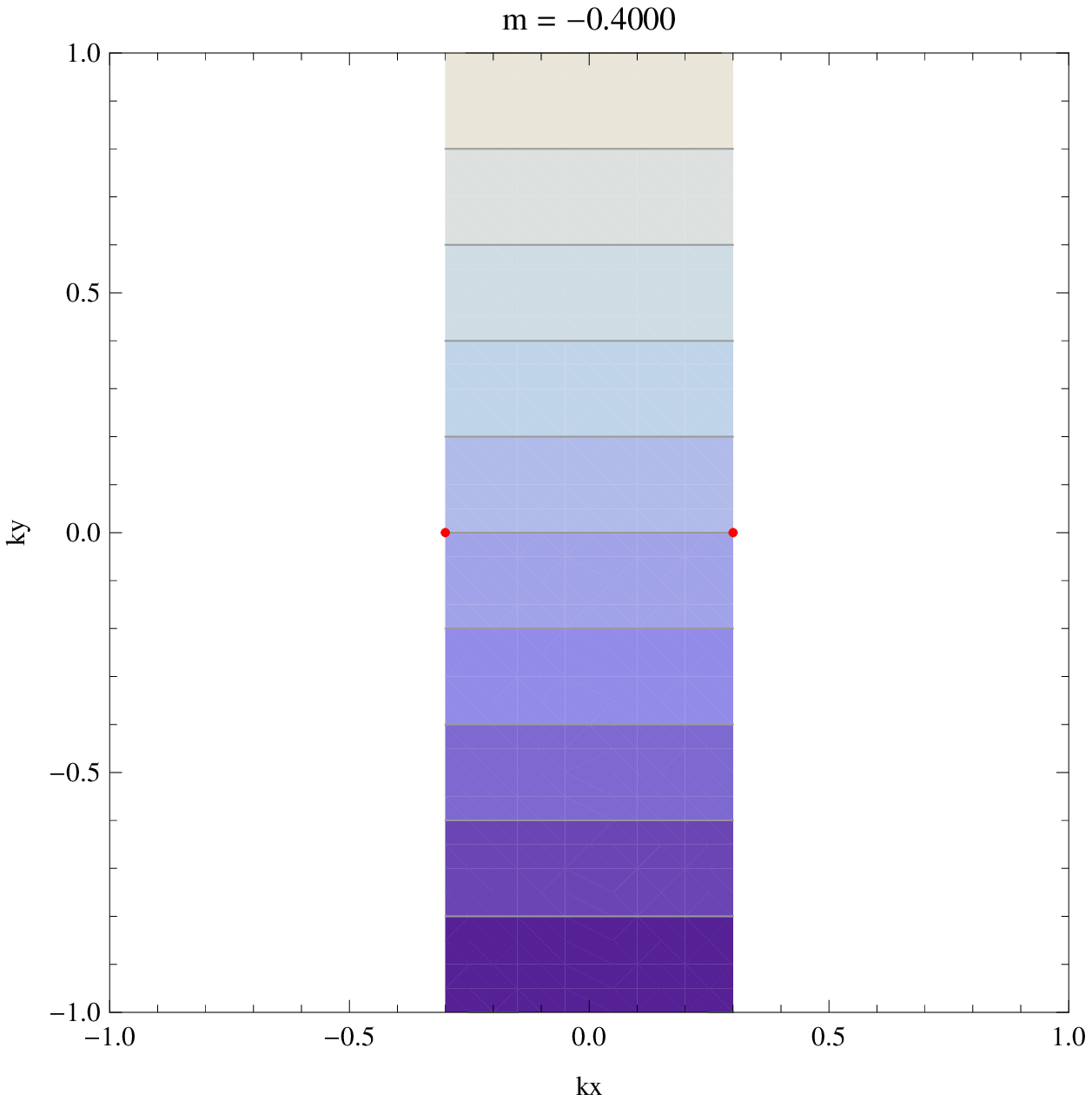}
  \includegraphics[width=5cm]{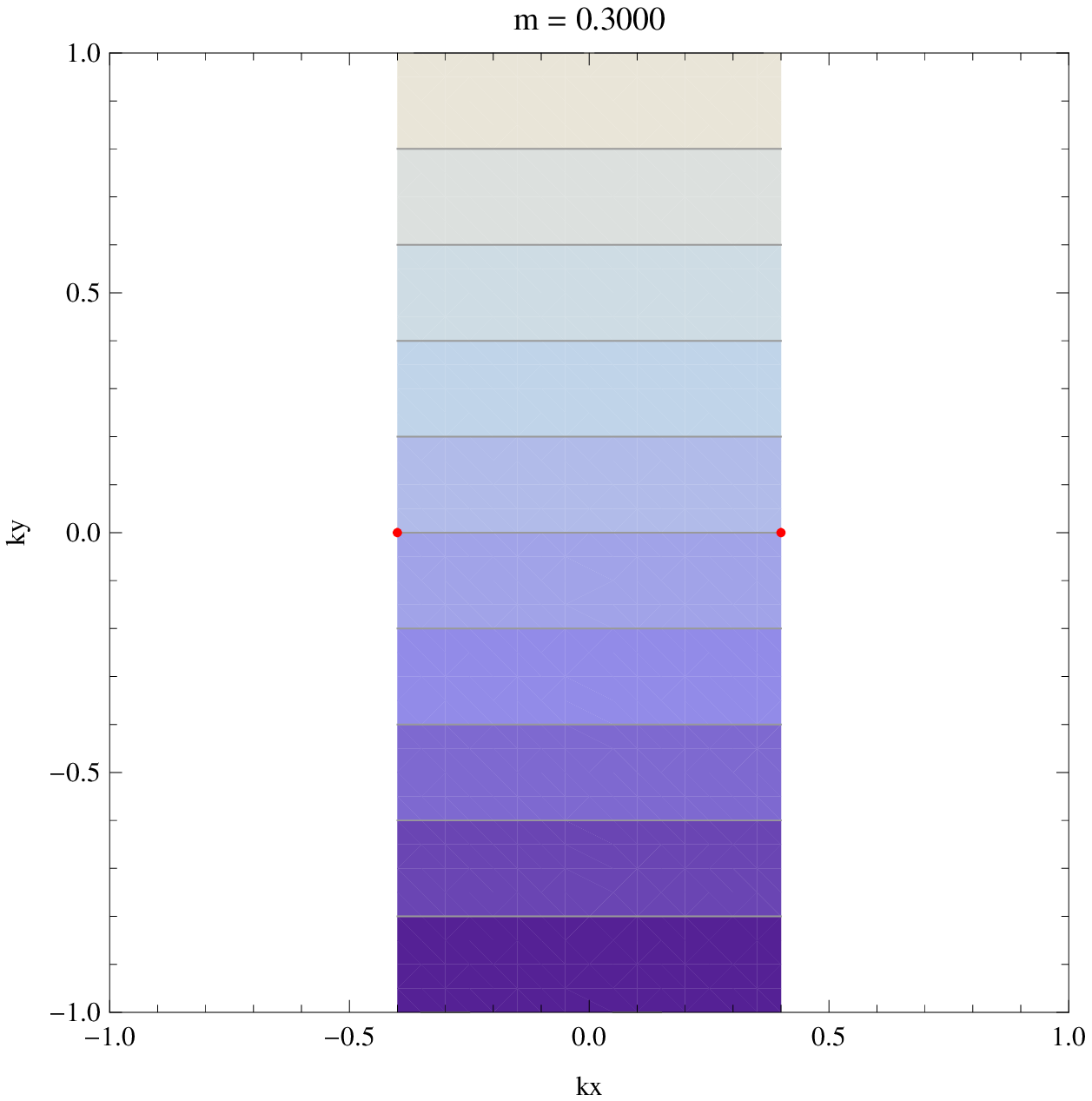}\\
  \caption{Contour plots of the dispersion relation of the surface states $\epsilon\lr{k_x, k_y}$ for the Weyl semimetal with a spatial momentum separation $2 b = 1.0$ ($b = 0.5$) between the Weyl nodes and different values of the Dirac mass $m$. Red points mark the projections of the bulk Weyl nodes ($k_x = \pm \sqrt{b^2 - m^2}$ if $|b| > |m|$, $k_x = 0$ otherwise, $k_y = 0$) onto the surface momentum space.}
  \label{fig:bs_spectrum_B}
\end{ltxfigure}

 Now we proceed as in Section (\ref{sec:bs_muA}) and represent the wave function $\Psi\lr{z}$ of the surface state as a linear combination of $\Psi_{+}\lr{z} = \lrc{\phi_{+}, \chi_{+}} \, e^{\lambda_{+} z}$ and $\Psi_{-}\lr{z} = \lrc{\phi_{-}, \chi_{-}} \, e^{\lambda_{-} z}$ with some coefficients $c_{+}$ and $c_{-}$. Matching $\Psi\lr{z = 0}$ to the wave function (\ref{IdealInsulatorEigenstate}) of an ideal insulator, we arrive at the following equations:
\begin{eqnarray}
\label{bs_B_eq1}
 c_{+} \phi_{+} + c_{-} \phi_{-} = \eta , \quad
 c_{+} \chi_{+} + c_{-} \chi_{-} = i \sigma_z \eta
 \, \Rightarrow \,
 c_{+} \lr{\chi_{+} - i \sigma_z \phi_{+}} + c_{-} \lr{\chi_{-} - i \sigma_z \phi_{-}} = 0 .
\end{eqnarray}
Taking into account the explicit form of $\chi_{\sigma}$ and $\phi_{\sigma}$ given by (\ref{phi_sigma_B}) and (\ref{chi_sigma_B}), we then arrive at the following compatibility condition for the above system of equations:
\begin{eqnarray}
\label{bs_B_eq3}
 \lr{\epsilon + i \lambda_{+} - \kappa \theta_{+} - i m}
 \lr{\lr{\epsilon - i \lambda_{-}} \theta_{-} - \bar{\kappa} + i \theta_{-} m}
 = 
 \lr{\epsilon + i \lambda_{-} - \kappa \theta_{-} - i m}
 \lr{\lr{\epsilon - i \lambda_{+}} \theta_{+} - \bar{\kappa} + i \theta_{+} m}
  ,
\end{eqnarray}
where we have denoted $\kappa = k_x - i k_y - b$. After some algebra and guesswork (guided by the numerical checks and the calculations of \cite{Goswami:12:1}) we solve these equations with respect to $\epsilon$ and arrive at the following results for the spectrum of the surface states:
 \begin{itemize}
 \item At $m > |b|$ no surface states exist.
 \item At $-|b| < m < |b|$ there are surface states for $|k_x| < \sqrt{b^2 - m^2}$, which have the simple dispersion relation $\epsilon = k_y$. In particular, the open line $|k_x| < \sqrt{b^2 - m^2}$, $k_y = 0$ at which $\epsilon = 0$ is the Fermi arc joining the projections of the bulk Weyl nodes at $k_x = \pm \sqrt{b^2 - m^2}$ \cite{Wan:11:1}.
 \item At $m < -|b|$ surface states exist for $|k_x| < m \sqrt{\frac{m^2}{b^2} - 1}$. The dispersion law is the anisotropic Dirac cone: the Fermi velocity in the $y$ direction $v_{F \, y} = \frac{\partial \epsilon}{\partial k_y}$ is always unity, and the Fermi velocity in the $x$ direction $v_{F \, x} = \frac{\partial \epsilon}{\partial k_x}$ is $v_{F \, x} = \sqrt{1 - \frac{b^2}{m^2}}$. The Dirac cone is particle-hole symmetric, that is, the energies of surface states always come in pairs $\pm \epsilon$.
 \item At $m = -|b|$ the surface states only exist at $k_x = 0$. Their dispersion relation is $\epsilon = \pm k_y$. Note the appearance of one more branch of the dispersion relation ($\epsilon = - k_y$) as compared to the dispersion relation at $-|b| < m < |b|$.
\end{itemize}

In order to illustrate these results, on Fig.~\ref{fig:bs_spectrum_B} we present the contour plots of the dispersion relation of the surface states $\epsilon\lr{k_x, k_y}$ for $b = 0.5$ and different values of $m$, both positive and negative. At $m < -|b|$ there are two opposite values of $\epsilon$ which correspond to the same $k_x$ and $k_y$, therefore we show the contour plots only for the branch with $\epsilon > 0$.

\section{Conclusions}
\label{sec:conclusions}

 In these notes we have given a general recipe for computing the spectrum of surface states for materials in which quasiparticle excitations are described by Dirac Hamiltonians of the form (\ref{DiracHamiltonianGeneral}) at low energies. We have also explicitly derived the spectra of low-energy surface states of 3D topological insulators and Weyl semimetals with broken parity and time-reversal symmetries.

 In the case when the parity is broken by the energy separation between the Weyl nodes we have found that the only effect of the energy separation is the reduction of the Fermi velocity in the Dirac cone dispersion relation of 2D surface states.

 The spectrum of surface states of massive Dirac Hamiltonian with broken time-reversal symmetry turned out to be more complicated. Physically, this Hamiltonian describes the magnetically doped 3D topological insulator. If the magnetic doping is sufficiently large, 3D topological insulator turns into a Weyl semimetal with momentum separation between the Weyl nodes. We have found that if the magnetic doping is not very large, its effect is to make the Dirac cone anisotropic by decreasing the Fermi velocity in the direction of the magnetization. At the same time, the range of momenta in which the surface states exist is shrunk to a strip of finite width, which is inversely proportional to the magnetization if the magnetization is small. As the magnetization becomes equal to the topological mass gap, this strip shrinks to a line which is perpendicular to the magnetization, and one branch of the dispersion relation disappears. As the magnetization is further increased, the bulk energy spectrum develops two separated Weyl nodes. Correspondingly, the surface states develop the Fermi arc and the dispersion relation becomes effectively one-dimensional: $\epsilon = k_y$. The surface states still exist in a finite strip with the width being equal to the separation between the bulk Weyl nodes. This picture of the evolution of the Dirac cone dispersion relation into the effectively one-dimensional Fermi arc dispersion relation might be useful for the identification of the critical magnetic doping of 3D topological insulators which leads to the emergence of a Weyl semimetal phase.

\subsection*{Acknowledgements}
This work was supported by the S. Kowalevskaja award from the Alexander von Humboldt foundation.



\end{document}